# Title: Response to Comment on (Novel two-dimensional porous graphitic carbon nitride C6N7 monolayer: A First-principle calculations [Appl. Phys. Lett. 2021, 119, 142102])

A. Bafekry,1 M. Faraji,2 M. M. Fadlallah,3 I. Abdolhosseini Sarsari,4 H. R. Jappor,5 S. Fazeli,6 and M. Ghergherehchi7

1Department of Radiation Application, Shahid Beheshti University, Tehran, Iran
2Micro and Nanotechnology Graduate Program, TOBB University of Economics and Technology, Sogutozu Caddesi No 43 Sogutozu, 06560 Ankara, Turkey
3Department of Physics, Faculty of Science, Benha University, 13518 Benha, Egypt
4Department of Physics, Isfahan University of Technology, Isfahan 84156-83111, Iran
5Department of Physics, College of Education for Pure Sciences, University of Babylon, Hilla, Iraq
6Department of Material Science and Engineering, Sharif University of Technology, PO Box 11155-9466, Tehran, Iran
7Department of Electrical and Computer Engineering, Sungkyunkwan University, 16419 Suwon, South Korea

Thank you for this comment. There are some typos in the values of the elastic parameters.
The values of $C_{11}$=258.6 GPa, $C_{22}$=290.8 GPa and $C_{12}$=70.73 GPa and $C_{13}=C_{23}$=2.49 GPa, $C_{33}$= 9.05 GPa, $C_{44}$= 25.86 GPa, $C_{55}$=2.90 GPa, and $C_{66}$= 3.08 GPa. The calculated Young's modulus is 362.9 GPa. Other elastic constants are related to the uniaxial elastic constant along the z-axis is found to be very small (due to the box of the unit cell), which shows that the interactions along the z-axis are minimized. It behaves as a perfect 2D structure.

The phonon bands exhibit two acoustic modes, a longitudinal one (LA) and an in-plane transverse one (TA), with linear dispersion ($\omega \propto q$) for $q \to 0$, and an out-of-plane flexure mode (ZA) with quadratic dispersion in the long-wavelength limit, which is a general feature of 2D membranes [1]. Flexural modes have been characterized in graphene [2,3], silicene [4,5], hexagonal boron nitride (h-BN) [6], MoS2 [7], and ultrathin silicon membranes [8,9]. Flexural modes play a role in elastic constants along the z-axis based on fundamental physics, and they can exist. However, the $C_{11}$ and $C_{12}$ values are generally the most critical elastic constants for monolayer materials [10]. The distribution of 6*6 elastic constants for selected materials and 3D (magenta), 2D (green), 1D (blue), and 0D (red) materials were shown in Fig. 7 of reference [11], which clearly shows the existence of $C_{13}$, $C_{23}$, and $C_{33}$ in the 2D systems [11]. As reported in Fig. 1e of our paper [12], ZA out-of-plane mode proves the existence of $C_{13}$, $C_{23}$, and $C_{33}$.

There are some ambiguities about their claim: 1-They can check the phonon dispersion of their structure to see ZA out-of-plane mode. 2-They report the uniaxial stress-strain responses in Fig 2., which is unrelated to our paper. For a more helpful understanding of the mechanical properties of the novel C6N7 monolayer, they can publish a paper. 3-They mentioned: "Using the DFT method and with assuming a thickness of 3.35 Å for the C6N7 monolayer based on graphene's thickness". Why did they choose this thickness while we know our C6N7 monolayer is flat without buckling? The distance of ZA out-of-plane movement of ions in C6N7 is different from Graphene.